\documentclass[aip,cha,twocolumn,amsmath,showpacs,superscriptaddress,reprint,a4paper]{revtex4-1}
\usepackage{rotating}
\usepackage{color}

\def\lb{\left}
\def\rb{\right}
\def\s{\sigma}
\def\nn{\nonumber}

\def\rtwo{\rho^{(2)}}

\def\pone{\phi^{(1)}}
\def\ptwo{\phi^{(2)}}
\def\szero{$S_0$}
\def\sone{$S_1$}
\def\stwo{$S_2$}
\def\D{\Delta}
\def\ER{Erd\H{o}s-R\'{e}nyi}
\def\kmax{{k_{\rm max}}}

\begin{document}

\title{Multi-Stage Complex Contagions}

\author{Sergey Melnik}
\affiliation{Oxford Centre for Industrial and Applied Mathematics, Mathematical Institute, University of Oxford, OX1 3LB, UK}
\affiliation{MACSI, Department of Mathematics \& Statistics, University of Limerick, Ireland}
\affiliation{CABDyN Complexity Centre, University of Oxford, Oxford, OX1 1HP, UK}
\author{Jonathan A. Ward}
\affiliation{Centre for Mathematics of Human Behaviour, Department of Mathematics \& Statistics, University of Reading, Reading, RG6 6AX, UK}
\author{James P. Gleeson}
\affiliation{MACSI, Department of Mathematics \& Statistics, University of Limerick, Ireland}
\author{Mason A. Porter}
\affiliation{Oxford Centre for Industrial and Applied Mathematics, Mathematical Institute, University of Oxford, OX1 3LB, UK}
\affiliation{CABDyN Complexity Centre, University of Oxford, Oxford, OX1 1HP, UK}



\begin{abstract}

The spread of ideas across a social network can be studied using complex contagion models, in which agents are activated by contact with multiple activated neighbors.  The investigation of complex contagions can provide crucial insights into social influence and behavior-adoption cascades on networks.  In this paper, we introduce a model of a multi-stage complex contagion on networks.  Agents at different stages --- which could, for example, represent differing levels of support for a social movement or differing levels of commitment to a certain product or idea --- exert different amounts of influence on their neighbors.  We demonstrate that the presence of even one additional stage introduces novel dynamical behavior, including interplay between multiple cascades, that cannot occur in single-stage contagion models.  We find that cascades --- and hence collective action --- can be driven not only by high-stage influencers but also by low-stage influencers.

\end{abstract}


\maketitle

\begin{quotation}
Studying models of cascades allows one to gain insights into a variety of processes ranging from the spread of fads and ideas in social networks to the appearance of cascading failures in infrastructure networks. To date, researchers have mostly considered single-stage cascade models wherein the propagation of a cascade is characterized by a single subpopulation of active agents,~\cite{Schelling73,Centola10,Watts02,Galstyan07,Gleeson08a} though some multi-stage models have been examined recently.\cite{debruyn08,Friedkin10} In the usual approach, it is assumed that all active agents exhibit the same amount of influence on their peers.  In reality, however, supporters of a cause can vary significantly in their desire and ability to recruit new members.  In this paper, we introduce a model of multi-stage cascading dynamics in which agents can exert different amounts of influence on their peers depending on the stage of their adoption (i.e., on the level of their commitment to a certain idea or product).  We investigate the dynamics of our multi-stage cascade model on various networks and observe an interplay between cascades --- e.g., one cascade driving the other one or vice versa --- that cannot be observed in single-stage cascade models.  We also provide an analytical method for solving the model that gives a good prediction for the cascade sizes on configuration-model networks.
\end{quotation}


\section{Introduction}

Social movements and other forms of collective action, which often arise spontaneously, require an ensemble of supporters with different levels of commitment.  Social influence and its potential to yield a critical mass of supporters can make a crucial difference as to whether or not movements succeed.~\cite{Oliver85,Oliver92,macy04,Macy91}  More generally, the effect that other people's opinions and actions have on the decisions that people make is a crucial sociological consideration,~\cite{Granovetter73,degroot74,Oliver85,Hedstrom00} and the impact that individual influence can have on the large-scale spread of rumors, fads, beliefs, and norms via social networks is of particular interest.~\cite{Watts02,Kempe03,Dodds05,Centola07a,Centola07b,Centola10,Young09,Martins09,Friedkin10,Friedkin11,McCullen11,Apt11,Krapivsky11,Valente96,Granovetter78,Friedkin01,Rogers03,fowler11,Lyons11,Snijders07,Christakis07,Goodreau09,Toole11,GonzalezBailon11,Centola11}  A closely related societal concern is that the mechanisms rooted in social interaction can give rise to financial crashes,~\cite{Devenow96} political revolutions,~\cite{Kuran89} successful technologies,~\cite{Venkatesh00} and cultural market sensations.~\cite{Salganik06}

The sudden changes in state exhibited in these examples are known as \emph{cascades}: Initially local behavior becomes widespread through collective action.  The perceived similarity between social and biological epidemics~\cite{Dodds05,Bettencourt06} has led to the use of the term \emph{contagion} for the spread of social influence.~\cite{Burt87}  Specifically, contagion refers to cases in which --- much like with a virus or a disease --- exposure to some source is enough to initiate propagation.  Importantly, social contagions need not just spread from one specific source to another.  In many situations, the chance of a node becoming \emph{active} (e.g., adopting a new technology or joining a political revolution) depends on several other people who are active --- and this is particularly true of people who are ``close'' or perceived as close in a social network.  Consequently, social contagions have been called \emph{complex contagions}~\cite{Centola07a,Centola07b,Centola10}.  Key investigations of complex contagions have included examinations of the diffusion of applications on the social networking site Facebook,~\cite{Onnela10} memes (short textual phrases) on news websites~\cite{Leskovec09} and other social media,~\cite{Simmons11,lerman12} information on blogs~\cite{Leskovec07b} and on the micro-blogging service Twitter,~\cite{Bakshy11} and voting in political elections.\cite{fowler12}

Although large data sets have the potential to help provide a better picture of social contagions, analyzing them without accompanying dynamical models offers little hope of distinguishing between underlying causes of individual behavior (social influence versus homophily versus covariates).~\cite{Shalizi11}  Statistical methods have been developed to approach the data side of this problem,~\cite{Aral09} but mathematical modelling is an underappreciated and crucial component of these efforts.  In particular, simple models make it possible to isolate effects (e.g., social contagion) and develop and test quantitative diagnostics that characterize macroscopic dynamics.

Early studies in the sociology of behavioral cascades considered threshold models of binary decisions.~\cite{Granovetter78}  In such models, agents can switch from an initially inactive state to an active state if a sufficient proportion of other agents are active.  These models capture two important features:~\cite{Oliver85} interdependence (an agent's behavior depends on the behavior of other agents) and heterogeneity (differences in behavior are reflected in the distribution of thresholds).  In recent studies,~\cite{Watts02,Gleeson07a,Galstyan07,deKerchove09} threshold models of social influence have been examined on networks in which the nodes correspond to agents and the edges between nodes indicate who influences whom.  In a network setting, models with simple threshold dynamics also capture fundamental mechanisms that can be related (at least by analogy) to a large range of phenomena that include failures in power grids and the transmission of infectious diseases.~\cite{LopezPintado08,Gleeson08a}  Tractable models of threshold dynamics have also sparked a great deal of interest in the physics and applied mathematics communities, because many results from graph theory, statistical physics, and dynamical systems can be applied directly in this setting.  For example, such techniques allow one to identify critical cascade thresholds,~\cite{Watts02} mean cascade sizes,~\cite{Gleeson08b} and the effects of seed size~\cite{Gleeson07a} and network topology.~\cite{Gleeson08a}

Motivated by the observation that not all opinions have equal weight, we introduce in this paper a model for \emph{multi-stage} complex contagions in which agents can adopt several different states with variable levels of influence.  Almost all existing models assume that active agents exert equal influence over their peers.  A notable exception is Ref.~\onlinecite{Friedkin10}, which recently considered networked agents with variable levels of social influence but used a very different framework from ours and also focused on different dynamics.  Another exception is Ref.~\onlinecite{debruyn08}, which considered multiple levels of social influence in the context of viral marketing but again used a very different approach.  As these papers illustrate, it is important to examine multi-stage social contagions in detail, as a binary description of agents' states can be woefully inadequate.  For example, it has been well-documented that supporters with varying level of commitment are crucial in social movements;~\cite{Oliver85,Oliver92} regular users of a product are more enthusiastic recommenders than casual users; supporters of a political party can vary significantly in their desire and ability to recruit new members; there are substantial behavioral differences between fans and fanatics of sports teams; and experience with free trial versions of software can greatly influence a user's decision to purchase a costly commercial version.

We show schematics of single-stage and two-stage complex contagion models in Fig.~\ref{fig_S012}.  To focus our discussion, we consider the simplest multi-stage progression, in which there are three possible states [see Fig.~\ref{fig_S012}(b)]: \emph{inactive} (no influence; unaware of an innovation or indifferent to a social movement), \emph{active} (low level of influence; testing an innovation or supporting a social movement), and \emph{hyper-active} (high level of influence; full adoption of an innovation or activists in a social movement).

\def\fwidth{1}
\begin{figure}[h]
\centering
\includegraphics[width=\fwidth\columnwidth]{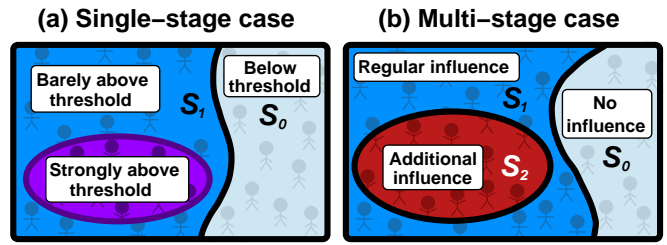}
\caption{(a) Schematic of a single-stage complex contagion.  All nodes can be either inactive (\szero) or active (\sone).  Nodes that are barely above the \sone-threshold have the same level of influence as nodes that are strongly above that threshold.  (b) Multi-stage complex contagion.  A subset of active nodes (called ``\sone-active") can become \emph{hyper-active} (called ``\stwo-active") and have additional influence.  Note that \stwo-active nodes are necessarily also \sone-active.}
\label{fig_S012}
\end{figure}

The rest of this paper is structured as follows.  In Secs.~\ref{Sec2} and~\ref{Sec3}, we define our model and illustrate some of its novel dynamics.  We demonstrate that it is possible for one cascade to drive the other one even in situations in which there would be no propagation (or only small amount of propagation) in the associated single-stage cascade model.  We then use synthetic networks and analytical approximations to further explore the model's dynamics, and we derive conditions for the appearance of cascades.  We include additional details in the appendices.


\section{Multi-Stage Model} \label{Sec2}

We consider situations in which there is an underlying social network, which we represent using an unweighted, undirected graph with $N$ nodes.  At any given point in time, there are three possible influence levels for each node: \emph{Inactive nodes} exert no influence on their neighbors and are said to be in state \szero, \emph{active nodes} exert some influence and are said to be in state \sone, and \emph{hyper-active nodes} exert additional influence and are said to be in state \stwo~[see Fig.~\ref{fig_S012}(b)].  Importantly, nodes in state \stwo~are a subset of nodes in state \sone, because nodes that are \stwo-active are necessarily \sone-active (i.e., we consider fanatics to be a specific type of fan), but \sone-active nodes need not be \stwo-active.  A natural generalization of the two-stage model includes further levels of influence such that an $S_n$-active agent is $S_i$-active for all $i \in \{1, \ldots, n\}$.

To specify the model, we also need to indicate precisely how nodes influence their neighbors and how the neighbors respond to such influence.  Accordingly, we define the influence \emph{response function} $F_i(m_1,m_2,k)$ as the probability that a degree-$k$ node (i.e., a node with $k$ neighbors) becomes $S_i$-active given that it has $m_1$ neighbors in state \sone~and $m_2$ neighbors in state \stwo.  This macroscopic description can be derived directly from a microscopic description of response functions of individual nodes.~\cite{LopezPintado08}  Formulating our model in such general terms allows us to capture a wide range of types of local interactions between nodes via the detailed form of the response functions $F_i$.~\cite{Gleeson08a}  In Sec.~\ref{Sec3}, we use a threshold model to illustrate the dynamics of multi-stage contagion models.


\section{Threshold Models} \label{Sec3}

We define the ``peer pressure'' $P = (m_1+\beta m_2)/k$ to be the total influence experienced by a degree-$k$ node from its $m_1$ neighbors in state \sone~and its $m_2$ neighbors in state \stwo, scaled by the node degree $k$.  Note that our definition of the states $S_i$ implies that a neighbor in state \stwo~contributes $1+\beta$ to the influence, so $\beta$ measures the \emph{bonus influence} exerted by \stwo-active nodes.  It is the presence of such bonus influence that distinguishes our multi-stage contagion model from single-stage models.

For threshold models of complex contagions, a node becomes $S_i$-active if the peer pressure $P$ is equal to or exceeds a certain threshold $R_i$ (which can be different for each node).  Therefore, the response function is $F_i(m_1,m_2,k)= C_i (P)$, where $C_i$ is the cumulative distribution of thresholds for state $S_i$ across network nodes.  In this paper, we focus mostly on uniform-threshold cases in which $S_i$-activation thresholds $R_i$ are the same for all nodes.  For uniform thresholds, the response functions are step functions:
\begin{align} \label{F_i_rel}
	F_i(m_1,m_2,k) =\lb\{\begin{array}{cl} 1,& {\rm if} \quad (m_1+\beta m_2)/k \ge R_i\\
                                               0,& {\rm otherwise} \end{array}\rb. \,.
\end{align}
We require that $R_2\ge R_1$ in order to satisfy $F_1(m_1,m_2,k)\ge F_2(m_1,m_2,k)$ and thereby guarantee that all \stwo-active nodes are also \sone-active.  When $\beta=0$, the \sone-state dynamics reduce to a single-stage contagion (i.e., with only one set of thresholds) because the \stwo-active nodes are indistinguishable from \sone~nodes.  As an initial condition, we select a small fraction $\phi^{(i)}$ of nodes to be initially $S_i$-active (and we note that they are never allowed to become less active). At each subsequent time step (of size $\D t = 1/N$), we update a randomly chosen node according to the threshold rules~\eqref{F_i_rel}.  This implies monotone dynamics---i.e., nodes can never become less active than they are currently.

In single-stage threshold models, it is usually the case that if a node needs a number $m$ of active neighbors to become active, then any subset of its neighbors with at least $m$ active nodes is sufficient to make it active.  In our multi-stage model, however, there is a heterogeneity in the subset of a node's neighbors needed for activation.  For example, the subset might consist of various possible combinations: 4 active neighbors, or 2 active and 1 hyper-active neighbor, or just 2 hyper-active neighbors.  Indeed, depending on a node's threshold values, various possible subsets of neighbors can make it active or even hyper-active. Moreover, for some parameters, a single hyper-active node might cause a cascade of activations even when a much larger number of active nodes that are spread throughout the network does not.

We now compare single-stage and multi-stage cascade dynamics and highlight situations that occur in multi-stage models that do not occur in the corresponding single-stage models.  Because these dynamics cannot arise if there are only two types of nodes (inactive and active), we see that the presence of hyper-active nodes can play a crucial role in driving cascades on networks.  We start by simulating these dynamics on the Facebook network of students at the University of Oklahoma (recorded in September 2005 as a single-time snapshot).~\cite{Traud08,datadump}


\subsection{Cascades Driven by High Influencers}

This example illustrates that a small amount of additional influence can trigger cascades.  This can model, for example, the role of a charismatic leader in a social movement or the potential effect of customer product reviews on retail websites such as Amazon.  In the context of our model, the presence of \stwo-active nodes triggers a cascade of \sone-active nodes that otherwise would not have occurred.  In particular, this effect arises specifically due to the extra influence attributed to \stwo-active nodes, which can be significantly above the \sone-activation threshold.

\def\fwidth{0.97}
\begin{figure}[h]
\flushright
\includegraphics[width=\fwidth\columnwidth]{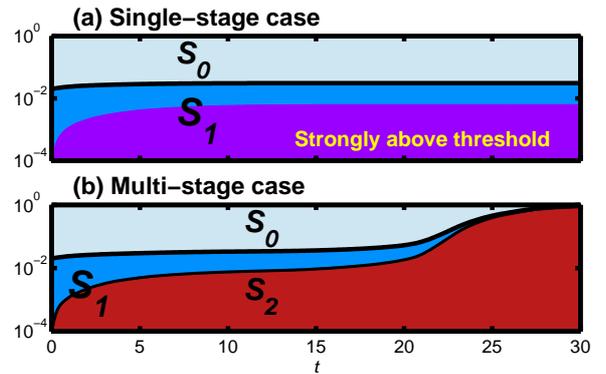}
\caption{Comparison of single-stage [panel (a)] and multi-stage [panel (b)] cascade dynamics on the Oklahoma Facebook network.~\cite{Traud08}  Time $t$ is on the horizontal axis, and we indicate the fractions of nodes in each state on the vertical axis.  Light blue, blue (and purple), and red regions represent \szero-, \sone-, and \stwo-active nodes, respectively.  In panel~(a), $\beta=0$; and in panel~(b), $\beta=0.5$.  The threshold parameter values are $R_1=0.15$ and $R_2=0.3$.  We use 348 \sone-active seed nodes (corresponding to $\pone \approx 0.02$) and 
zero \stwo-active seeds ($\ptwo=0$).  There is no cascade in panel~(a), but some of the nodes (colored in purple) are well above the \sone-threshold.  The bonus influence of \stwo-active nodes drives a cascade in panel~(b).}
\label{S2drS1}
\end{figure}

In Fig.~\ref{S2drS1}, we compare the dynamics of a single-stage contagion [panel (a); $\beta = 0$] and a multi-stage contagion [panel (b); $\beta = 0.5$].  Both cases use the response function defined in Eq.~\eqref{F_i_rel} with parameter values $R_1=0.15$ and $R_2=0.3$.  In both cases, we show an average over 100 realizations with the same initially active nodes.~\cite{Note1}  In Fig.~\ref{S2drS1}(a), \stwo-active nodes have no influence on the activation of \sone-active nodes (because $\beta = 0$) and there is no cascade: the fraction of \sone~nodes remains small.  Note, however, that some nodes are well above the activation threshold (purple region).  Specifically, these nodes surpass the threshold $R_2=0.3$ but have no additional influence.  In Fig.~\ref{S2drS1}(b), \stwo-active nodes that surpass the higher threshold $R_2=0.3$ have 1.5 times as much influence as \sone-active nodes.  This additional influence is enough to trigger a global (system-wide) cascade.


\subsection{Cascades Driven by Low Influencers}

This example illustrates that a small number of additional \emph{low-level} influencers that are \sone-active can also trigger a cascade.  This could represent situations in which a company gives out free trials of a product to potential customers with the aim of boosting sales of the full product.  Once again, we use the Oklahoma Facebook network to compare two example situations.  In the first, we set $R_1=R_2=0.2$; in the second, we set $R_1=0.15$ and $R_2=0.2$.  The first case is essentially a single-stage process because the \sone~dynamics are slaved to the \stwo~dynamics.  The other parameters are the same in both simulations: $\beta=0.3$ and $\phi^{(1)}=\phi^{(2)} \approx 0.02$.  The response function is again given by Eq.~\eqref{F_i_rel}.  We again average over 100 realizations using the same set of initially active nodes for both cases.~\cite{Note2}

There is no cascade in Fig.~\ref{S1drS2}(a), in which the activation thresholds are equal.  In Fig.~\ref{S1drS2}(b), however, the \sone-activation threshold is slightly lower.  This results in a small number of additional \sone-active nodes, which is enough to trigger a cascade.

\def\fwidth{0.97}
\begin{figure}[h]
\flushright
\includegraphics[width=\fwidth\columnwidth]{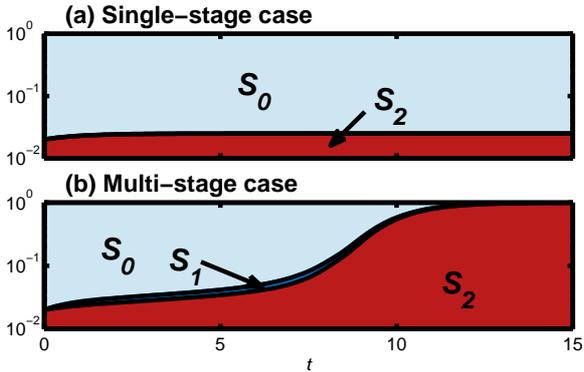}
\caption{Comparison of single-stage [panel (a)] and multi-stage [panel (b)] cascade dynamics on the Oklahoma Facebook network.~\cite{Traud08}  As in Fig.~\ref{S2drS1}, the time $t$ is on the horizontal axis, and we indicate the fractions of nodes in each state on the vertical axis.  Light blue, blue, and red regions represent \szero-, \sone-, and \stwo-active nodes, respectively.  We use 348 \stwo-active seed nodes (corresponding to $\ptwo=\pone \approx 0.02$) and $\beta=0.3$.  In panel~(a), $R_1=R_2=0.2$, so the \sone~dynamics are slaved to the \stwo~dynamics.  A small change in the threshold parameter ($R_1=0.15$) in panel~(b) yields a small number of additional \sone-active nodes, which are nevertheless enough to trigger a cascade.}
\label{S1drS2}
\end{figure}


\section{Synthetic Networks}

To better understand the cascade dynamics, it is instructive to consider well-chosen synthetic networks that make it possible to control the final cascade sizes.

We design synthetic scenarios in which some nodes do not become active in the single-stage case, but the cascade in the multi-stage case occupies the entire system.  We construct ensembles of random networks consisting of nodes of degrees 4 and 24 in proportion 1:1 with positive degree-degree correlations.  (We dub such networks ``(4,24)-correlated random networks"; degree assortativity is positive when degree-degree correlations are positive.) We then use the response function defined in Eq.~\eqref{F_i_rel} that scales the activation thresholds by node degree.  It is thereby harder for high-degree nodes to become active because they need a larger number of active neighbors (most of which are also of high degree).  We construct the network using the method described in Ref.~\onlinecite{Dorogovtsev03}, where the assortativity is captured by the joint distribution $P(k,k')$, which gives the probability that a randomly chosen edge connects a node of degree $k$ to a node of degree $k'$.  For our example, we choose $P(k,k')$ to be a (symmetric) matrix whose non-zero entries satisfy $P(4,4)/P(4,24)=3$ and $P(24,24)/P(24,4)=23$.  This gives a network with a specified amount of (positive) assortativity.  An additional consequence of the multi-stage dynamics on such networks is that a trajectory tends to spend a significant amount of time near a value that is not the final steady state.

In the first example, which we illustrate in Fig.~\ref{S2drS1a_asynch}, the extra influence exerted by \stwo-active nodes is needed to drive a system-wide cascade.  To demonstrate this effect, we compare the dynamics resulting from a single-stage case, which is captured by the \sone-dynamics of the multi-stage case with an upper threshold $R_2=\infty$, with those resulting from a (true) multi-stage case with $R_2 = 0.7$.  (See the caption of Fig.~\ref{S2drS1a_asynch} for the values of the other parameters.) In Fig.~\ref{S2drS1a_asynch}, we show the results of numerical simulations using symbols and analytical results given by Eqs.~\eqref{rhoki}--\eqref{qki} using curves.  The analytical results qualitatively reproduce the numerical behavior; in some cases, we also observe good quantitative agreement.  However, as discussed in Appendix~\ref{appB}, some novel effects arising in the multi-stage model are not captured by our analytical approximation and can potentially lead to incorrect estimates.  We show the aggregate fractions of \sone- and \stwo-active nodes for the single-stage case in Fig.~\ref{S2drS1a_asynch}(a) and for the multi-stage case in Fig.~\ref{S2drS1a_asynch}(b).  We show the separate temporal evolutions for the two degree classes in Figs.~\ref{S2drS1a_asynch}(c,d), and we show and the temporal evolution for nodes in each degree class that are \sone- but not \stwo-active in Fig.~\ref{S2drS1a_asynch}(e).

In the single-stage case, only the low-degree nodes ultimately become \sone-active, which results in the aggregate active fraction of $0.5$ illustrated in Fig.~\ref{S2drS1a_asynch}(a).  Observe that none of the high-degree nodes become \sone-active [see Fig.~\ref{S2drS1a_asynch}(c)] and that no nodes become \stwo-active because $R_2=\infty$.  In the multi-stage case, which we show in Fig.~\ref{S2drS1a_asynch}(b), the low-degree nodes that were significantly above the \sone-activation threshold in the single-stage case eventually become \stwo-active and consequently exert more influence on their neighbors.  As we show in Fig.~\ref{S2drS1a_asynch}(d), this initiates a gradual increase in the number of \sone-active high-degree nodes (starting around $t=10$) until there are sufficiently many such nodes to trigger a delayed rapid transition or a secondary cascade (around $t=13$) in which all nodes become \stwo~active.

\def\fwidth{1}
\begin{figure}[h]
\centering
\includegraphics[width=\fwidth\columnwidth]{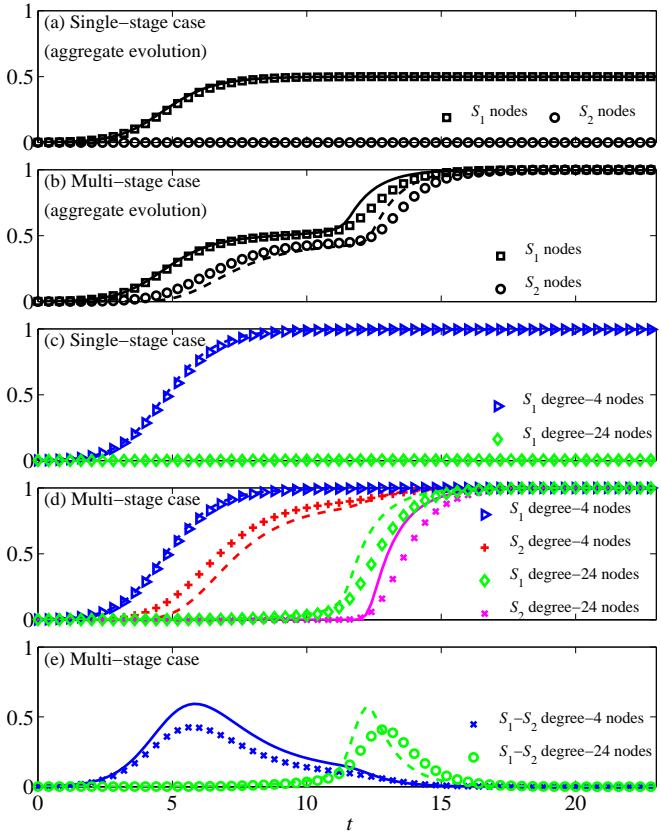}
\caption{Demonstration of dynamics when an \stwo-cascade drives an \sone-cascade.  Panels (a) and (b) show the aggregate fractions of \sone- and \stwo-active nodes, panels (c) and (d) show these fractions for each degree class separately, and panel (e) shows the fractions of nodes in each degree class that are \sone-active but not \stwo-active.  We show the numerical results (averaged over 100 realizations) using symbols and the analytical results given by Eqs.~\eqref{rhoki}--\eqref{qki} using curves.  The timescales are independent of network size $N$, which we take to be $N=10^4$.  The values of the other parameters are $R_1=0.2$, $R_2=0.7$, $\beta=0.45$, $\pone = 10^{-3}$, and $\ptwo = 0$.  (We choose seed nodes uniformly at random.)  We use an upper threshold of $R_2=\infty$ to model the single-stage case.}
\label{S2drS1a_asynch}
\end{figure}

In Fig.~\ref{S2drS1a_asynch}(e), we show for each degree class the temporal evolution of the fraction of the \sone-active nodes that are not \stwo-active.  The number of such nodes is given by the difference between \sone-active and \stwo-active nodes.  The peaks in Fig.~\ref{S2drS1a_asynch}(e) imply that nodes in each degree class first become \sone~active and that there is some delay for their \stwo-activation.

Now imagine that only the \sone-dynamics in Fig.~\ref{S2drS1a_asynch}(b) can be observed.  For example, suppose that a publisher knows which nodes purchased a particular book but has no idea about the level of excitement of any of the nodes.  In this case, at time $t=8$, the publisher might mistakenly conclude that the product is not going to be sold anymore and could perhaps discontinue it.  However, the heterogeneity of excitement and influence levels among book purchasers suggests instead that the publisher should continue printing more copies of the book.  In this toy example, there would be a sharp rise in sales at time $t=13$ caused by the \stwo-active agents (who, e.g., could represent customers who write book reviews on Amazon).

\def\fwidth{1}
\begin{figure}[h]
\centering
\includegraphics[width=\fwidth\columnwidth]{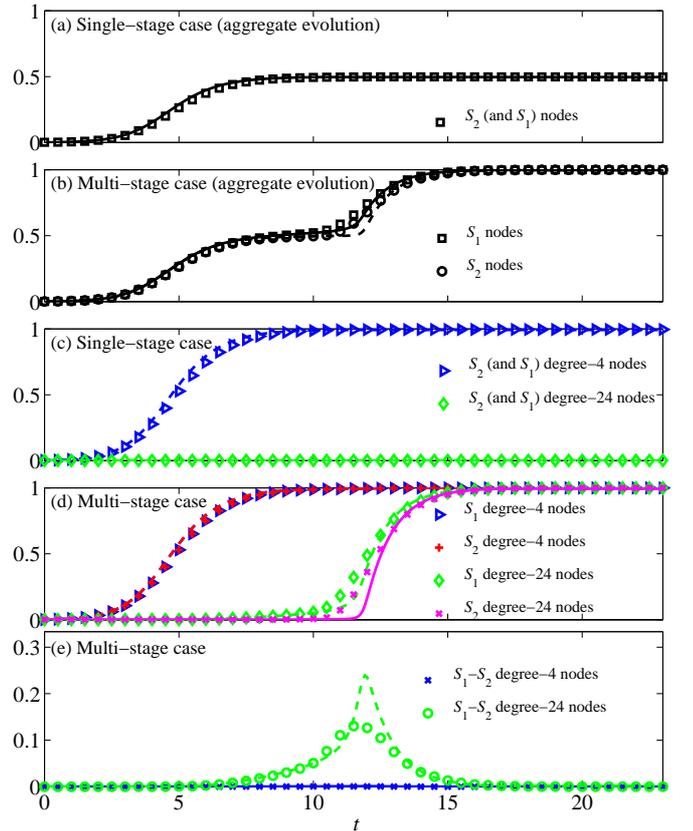}
\caption{Demonstration of dynamics when an \sone-cascade drives an \stwo-cascade.  Panels (a) and (b) show the aggregate fractions of \sone- and \stwo-active nodes, panels (c) and (d) show these fractions for each degree class separately, and panel (e) shows the fractions of nodes in each degree class that are \sone-active but not \stwo-active.  We show the numerical results (averaged over 100 realizations) using symbols and the analytical results given by Eqs.~\eqref{rhoki}--\eqref{qki} using curves.  The timescales are independent of network size $N$, which we take to be $N=10^4$.  We use the values $R_1=R_2=0.35$ for the single-stage case [panels (a) and (c)] and the values $R_1=0.22$ and $R_2=0.35$ for the multi-stage case [panels (b), (d), and (e)].  The values of the other parameters are $\beta=0.45$ and $\pone = \ptwo = 10^{-3}$ (where we choose the seed nodes uniformly at random).}
\label{S1drS2a_asynch}
\end{figure}

We now consider another example, illustrated in Fig.~\ref{S1drS2a_asynch}, in which the addition of an \sone-active state enhances the propagation of an \stwo-cascade, which (in this case) eventually reaches the entire network.  In Fig.~\ref{S1drS2a_asynch}(a), we illustrate the case in which $R_1=R_2=0.35$.  This is effectively a single-stage scenario because the \stwo-active nodes are also \sone~active.  (See the caption of Fig.~\ref{S1drS2a_asynch} for the values of all parameters.)  Similar to the previous example, all of the low-degree nodes become \sone-active---but now they are also all \stwo~active.  None of the high-degree nodes become active [see Fig.~\ref{S1drS2a_asynch}(c)].  In order to obtain a (true) multi-stage scenario, we reduce the $R_1$ threshold to $0.22$ and show the resulting aggregate dynamics in Fig.~\ref{S1drS2a_asynch}(b).  The dynamics are initially qualitatively similar to the single-stage case of Fig.~\ref{S1drS2a_asynch}(a).  However, after some time passes, there is an \sone-activation surge of high-degree nodes due to the lower \sone~threshold.  They subsequently drive an \stwo-activation cascade [see Fig.~\ref{S1drS2a_asynch}(d)].

In Fig.~\ref{S1drS2a_asynch}(e), we show the temporal evolution of the fraction of the \sone-active nodes that are not \stwo-active.  This quantity exhibits a peak for the high-degree nodes but remains at zero for low-degree nodes.  This indicates that high-degree nodes become \stwo-active some time after becoming \sone-active, whereas low-degree nodes become \stwo-active and \sone-active simultaneously.


\section{Analysis}

We now present an analytical approximation for the temporal evolution of the fraction of active nodes in our multi-stage model.  The method that we employ is based on pairwise interactions between nodes~\cite{Gleeson07a,Gleeson08a} and entails two requirements: (i) for any fixed $k$, the response functions $F_i$ must be non-decreasing functions of both $m_1$ and $m_2$; and (ii) $F_i(m_1,m_2,k)\ge F_{i+1}(m_1,m_2,k)$.  Condition (i) reflects the effect of positive externalities: when a node has more active neighbors, it is more likely to become $S_i$-active.  Condition (ii) follows from the fact that the number of hyper-active nodes should not exceed the number of all active nodes in the system.  The situations that we illustrate in this paper satisfy these conditions, which usually tend to be sensible assumptions when studying social influence.~\cite{notehipster}

One computes the density $\rho_k^{(i)}(t)$ of degree-$k$ nodes that are $S_i$-active at time $t$ by solving the set of ordinary differential equations
\begin{align}
	\label{rhoki} \dot \rho_k^{(i)}(t) &= H_k^{(i)}\lb(\rho_k^{(i)}(t),Q^{(1)}(t),Q^{(2)}(t)\rb)\,,\\
	\label{qki} \dot Q^{(i)}(t) &= G^{(i)}\lb(Q^{(1)}(t),Q^{(2)}(t)\rb)\,,
\end{align}
where $Q^{(i)}$ is a vector of auxiliary variables.  We present the functionals $H_k^{(i)}$ and $G^{(i)}$ (each of which depends on $F_i$ and the network topology) and the derivation of Eqs.~(\ref{rhoki})--(\ref{qki}) in Appendix~\ref{appA}.  The fraction $\phi^{(i)}$ of nodes that are initially $S_i$-active specifies the initial conditions $\rho_k^{(i)}(0) = \phi^{(i)}$ and $Q^{(i)}(0) =\lb[\phi^{(i)},\ldots,\phi^{(i)}\rb]$.  The aggregate fraction of nodes that are $S_i$-active at time $t$ is $\rho^{(i)}(t) = \sum_{k=0}^{\kmax} P_k \rho_k^{(i)}(t)$, where $P_k$ is the degree distribution of the network and $\kmax$ is the maximum node degree in the network.


\section{Final State and Temporal Evolution of Cascades}

In Fig.~\ref{figPk45}, we compare the analytical predictions of Eqs.~\eqref{rhoki}--\eqref{qki} with numerical simulations of both the final values and the temporal evolution of active fractions of nodes.  This figure uses two example network ensembles: (i) \ER~random graphs (i.e., graphs in which each pair of nodes is connected by an edge with equal, independent probability) with mean degree $z=5$; and (ii) uncorrelated random graphs (i.e., random graphs whose joint degree-degree distribution can be expressed in terms of the degree distribution as $P(k,k') = kP_k k' P_{k'}/z^2$) consisting of degree-4 and degree-5 nodes in proportion 1:2.  We dub the latter graphs ``(4,5)-uncorrelated random networks". In this example, we use the response functions $F_i(m_1,m_2,1)$ to obtain a threshold model in which the threshold conditions are based on the number --- rather than on the fraction --- of active neighbors. (This variant of a threshold model was used, for example, in Ref.~\onlinecite{Galstyan07} to investigate cascading dynamics in modular networks.) In this situation, the peer pressure experienced by a node from its neighbors is not scaled by the node's degree $k$, so the number of active neighbors required to activate a node is independent of degree. Additionally, an active node influences all of its neighbors equally (regardless of their degrees), and high-degree nodes can thus become active more easily.

As one can see in Fig.~\ref{figPk45}, Eqs.~\eqref{rhoki}--\eqref{qki} correctly predict the final fractions of active nodes in all cases and are in good agreement with the numerically computed temporal evolution.  (Note that we do not plot the final fraction of \sone-active nodes, as it equals 1 in each of the examples illustrated in this figure.)  Observe that the agreement between theory and simulation in Fig.~\ref{figPk45} is better than that in Figs.~\ref{S2drS1a_asynch} and~\ref{S1drS2a_asynch} because the effects described in Appendix~\ref{appB} have little impact in this situation.

\def\fwidth{0.95}
\begin{figure}[h]
\centering
\includegraphics[width=\fwidth\columnwidth]{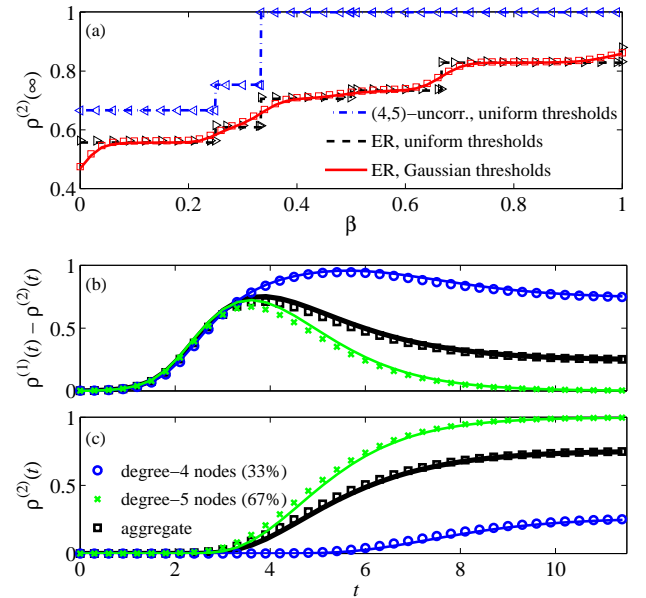}
\caption{Comparison of numerical computations (symbols) with analytical predictions of Eqs.~\eqref{rhoki}--\eqref{qki} (curves) for (a) the final fractions of \stwo-active nodes as functions of the bonus influence $\beta$ for ensembles of (4,5)-uncorrelated random networks (blue dash-dotted curve) and \ER~random graphs with mean degree $z=5$ (black dashed curve) and (b,c) the temporal evolution of active fraction of nodes in each degree class in (4,5)-uncorrelated random networks for $\beta=0.25$.  In panel (a), we use the response functions of Eq.~\eqref{F_i_rel} with $k\equiv 1$ and uniform thresholds $R_1 = 1$ and $R_2 = 5$. For the ER graphs, we also use a solid red curve to show the case in which the $R_2$ thresholds are Gaussian-distributed with mean $\mu=5$ and standard deviation $\s=0.1$.  The total number of nodes in each network is $N=10^4$.  For the numerical simulations, we initially \sone-activate a fraction $\pone = 10^{-3}$ of nodes chosen uniformly at random, and we average over 100 realizations of networks and initial conditions.}
\label{figPk45}
\end{figure}

In Fig.~\ref{figPk45}(a), the blue dash-dotted curve corresponding to the (4,5)-uncorrelated random network has steps at $\beta=1/4$ and $\beta=1/3$.  These steps arise as follows: When $\beta<1/4$, degree-4 nodes can never experience enough peer pressure from their neighbors to overcome the threshold $R_2=5$ to become \stwo-active (see Eq.~\eqref{F_i_rel} with $k\equiv 1$).  Therefore, the final fraction of \stwo-active nodes $\rtwo(\infty)$ is given by the fraction of degree-5 nodes in the network (which is equal to $2/3$ in this example).  When $\beta \in [1/4,1/3)$, degree-4 nodes become  \stwo-active only when all of their neighbors are \stwo-active.  Consequently, a finite fraction (about $0.26$) of degree-4 nodes becomes \stwo-active, yielding the aggregate value of $\rtwo(\infty) \approx 0.75$.  We show the temporal evolution of the active fraction of nodes for each degree class for this case in Fig.~\ref{figPk45}(b,c).  Finally, for $\beta\ge 1/3$, degree-4 nodes become \stwo-active if all of their neighbors are \sone-active and any 3 of them are \stwo-active.  In this situation, all degree-4 nodes become \stwo-active by the end of the cascade.


\section{Cascade Condition and Bifurcation Analysis}

In Fig.~\ref{figcc}, we illustrate the relationship between the final fraction of \sone-active nodes $\rho^{(1)}_{\infty}$, the bonus influence $\beta$, and the mean degree $z$ for \ER~random graphs.  Darker colors indicate larger final activation fractions.  The final fraction of \stwo-active nodes is qualitatively similar.  Note that when $\beta=0$ (i.e., single-stage dynamics), there is no cascade; as $\beta$ increases from $0$ (for fixed values of $z$), we observe transitions that are qualitatively similar to those of Fig.~\ref{figPk45}(a).

\def\fwidth{0.97}
\begin{figure}[t]
\centering
\includegraphics[width=\fwidth\columnwidth]{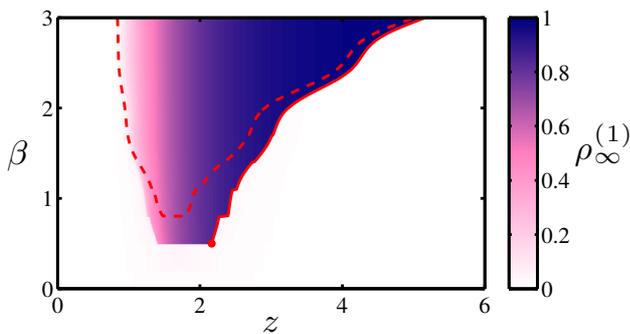}
\caption{Two-parameter bifurcation diagram for $\rho^{(1)}_\infty$ (whose value is indicated by color) calculated from Eqs.~\eqref{rhoki}--\eqref{qki} for \ER~random graphs.  The mean degree $z$ is on the horizontal axis, and the bonus influence $\beta$ is on the vertical axis.  The first threshold is $R_1=0.3$, and the second threshold $R_2$ is Gaussian-distributed with mean $\mu=0.8$ and standard deviation $\sigma=0.2$.  The initial seed fractions are $\phi^{(1)}=2\times10^{-3}$ and $\phi^{(2)}=0$.  The dashed curve gives the boundary of the cascade condition, and the solid curve is a numerical continuation of the saddle-node bifurcation.}
\label{figcc}
\end{figure}

A bifurcation analysis gives an analytical estimate of the boundary of the region in which cascades occur.  In analogy to the methods developed for single-stage models,~\cite{Gleeson07a,Gleeson08a} we derive a cascade condition from Eqs.~\eqref{rhoki}--\eqref{qki}.  (We present full details in Appendix~\ref{appC}.) Briefly, we compute the zero eigenvalues of the Jacobian matrix of $G$ evaluated at $Q^{(1)}=Q^{(2)}=0$.  This yields a closed-form expression that approximates the boundary of the cascade region.  This approximation, which is crude but given by a closed-form expression (see Appendix~\ref{appC}), yields the dashed curve in Fig.~\ref{figcc}.

For small values of $z$, we find a continuous transition from small to large values of $\rho^{(1)}_\infty$, which reflects the distribution of small component sizes of the graph (for further details, see Ref.~\onlinecite{Gleeson07a}). In contrast, for larger values of $z$, we find a discontinuous transition. This jump arises from a saddle-node bifurcation that occurs as $z$ is increased while $\beta$ is held fixed.  This bifurcation can be followed numerically by solving $G=0$ and finding zero eigenvalues of the Jacobian evaluated at the corresponding equilibria (see the solid curve in Fig.~\ref{figcc}). We provide full details in Appendix~\ref{appC}.

As in the Watts single-stage threshold model,~\cite{Watts02} numerical simulations using parameter values close to a saddle-node bifurcation are very sensitive to the choice of seed nodes.  A recent (and somewhat controversial) empirical study~\cite{goel11} has suggested that real cascades are extremely rare events.  This is consistent with the dynamics that occur near the saddle-node bifurcation in the present example: a few specific choices of initial seeds produce large-scale cascades but most choices do not.

\def\fwidth{0.97}
\begin{figure}[t]
\centering
\includegraphics[width=\fwidth\columnwidth]{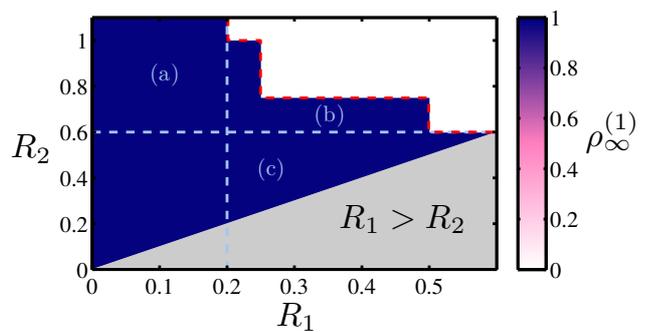}
\caption{Two-parameter bifurcation diagram for $\rho^{(1)}_\infty$ (whose value is indicated by color) calculated from Eqs.~\eqref{rhoki}--\eqref{qki} for \ER~random graphs with mean degree $z=4$. We plot the uniform thresholds $R_1$ and $R_2$ on the horizontal and vertical axes, respectively. The bonus influence is $\beta=2$, and the initial seed fractions are $\phi^{(1)}=1\times10^{-4}$ and $\phi^{(2)}=0$. The dashed red line gives the boundary of the cascade condition. The labeled regions (which are separated by dashed white lines) indicate cascades that are driven by (a) low influencers; (b) low and high influencers; and (c) high influencers. See the description in the text.}
\label{figR1R2}
\end{figure}

In Fig.~\ref{figR1R2}, we illustrate the relationship between the final fraction of \sone-active nodes $\rho^{(1)}_{\infty}$ and the uniform thresholds $R_1$ and $R_2$ for \ER~random graphs with mean degree $z=4$. The bonus influence is $\beta=2$, and the initial seed fractions are $\phi^{(1)}=1\times10^{-4}$ and $\phi^{(2)}=0$. As in Fig.~\ref{figcc}, darker colors indicate larger final activation fractions and the cascade condition is indicated by the dashed red line. The corresponding plot of the final fraction of \stwo-active nodes as a function of $R_1$ and $R_2$ is qualitatively similar to Fig.~\ref{figR1R2} except for $R_2>1$, for which there are no cascades of high influencers because $\phi^{(2)}=0$. The region $R_1>R_2$ is forbidden by our definitions of $R_1$ and $R_2$. Recall that (i) cascades are driven by high influencers if there is a cascade for some $\beta>0$ when there is none at $\beta=0$; and (ii) cascades are driven by low influencers if there is a cascade for some $R_1<R_2$ but there is none at $R_1=R_2$.  We can identify regions in Fig.~\ref{figR1R2} corresponding to cascades driven by (a) low influencers; (b) low and high influencers; and (c) high influencers.  When $\beta=0$ and all other parameters are held constant, the value $R_1\approx0.2$ marks a boundary between a region without cascades (to the right) and a region in which cascades are possible (to the left).  Thus, the cascades in Fig.~\ref{figR1R2} that occur to the right of $R_1\approx0.2$ are driven by high influencers. Similarly, cascades that occur above $R_2 \approx 0.6$ are driven by low influencers, because increasing $R_1$ eventually results in no cascades.


\section{Conclusions}

Social movements and other forms of collective action require an ensemble of supporters with different levels of commitment, and social influence can make a crucial difference as to whether or not they succeed.  This motivates the development of analytically tractable complex contagion models with multiple stages in which different agents have different levels of influence.

In the present paper, we have introduced and analyzed such a model, in which we define the level of influence on a node from its neighbors using a general function of the node's degree and the state of its nearest neighbors.  We illustrated that this model can exhibit interesting dynamics that are not possible with single-stage cascade models.  This includes, in particular, the interplay between the cascades of fans (active nodes) and fanatics (hyper-active nodes), in which one cascade can drive the other and vice versa.  Our model and our analytical results can be generalized to multi-stage cases with any finite number of active states.  The model can also be developed further to allow one to distinguish the level of a node's commitment and the influence it has on its neighbors.

Different levels of commitment and influence have well-documented importance on social movements, product advertising, and other sociological, political, and economic situations.  However, mere observation and data analysis of complex social dynamics make it difficult to discern the relative importance of social influence, homophily, and covariates on observations.~\cite{Shalizi11}  It is therefore imperative to develop new mathematical models to tackle this challenging situation, and we hope that the model we have introduced in this paper will prove beneficial for such efforts.


\section*{Acknowledgements}

This work was funded in part by the Irish Research Council co-funded by Marie Curie Actions under FP7 (INSPIRE fellowship, S.M.), the Engineering and Physical Sciences Research Council (MOLTEN, EP/I016058/1, J.A.W.), Science Foundation Ireland (11/PI/1026, J.P.G.), the James S. McDonnell Foundation (award \#220020177, M.A.P.), and the European Commission FET-Proactive project PLEXMATH (FP7-ICT-2011-8, grant number 317614, M.A.P. and J.P.G.). Work by S.M. and M.A.P. was carried out in part at the Statistical and Applied Mathematical Sciences Institute in the Research Triangle Park, North Carolina.  We thank Adam D'Angelo and Facebook for providing the Facebook data used in this study.  We thank James Moody, Jukka-Pekka Onnela, Noah E. Friedkin, Mariano Beguerisse D\'{\i}az, and Sandra Gonz\'{a}lez-Bail\'{o}n for reading and commenting on a draft of the paper.  We also thank Peter Grindrod and Ben Williamson for useful comments and a referee for suggesting Fig.~\ref{figR1R2}.


\appendix
\section{Analytical Approximation} \label{appA}

In this section, we derive approximate analytical results for the time-dependence of the density of active nodes in our multi-stage model.  The derivation is based on pairwise interactions between nodes and builds on the method described in Refs.~\onlinecite{Gleeson07a,Gleeson08a}.  We first consider the synchronous updating case in which the states of all $N$ network nodes are updated at each discrete time step.  We then extend the results to situations in which only the states of a certain fraction $\tau$ of nodes (chosen uniformly at random) are updated.  Thus, the value $\tau=1$ corresponds to synchronous updating of all nodes and $\tau=1/N$ corresponds to completely asynchronous updating in which a single (randomly chosen) node is updated at each time step.  For our model, both types of updating lead to the same final density values, but the transient behavior can be different.

We focus on a class of undirected, unweighted random networks defined by the joint degree-degree distribution $P(k,k')$, which gives the probability that a randomly chosen edge connects a node of degree $k$ to a node of degree $k'$ (with everything else selected uniformly at random).  This class of random networks reduces to configuration-model networks defined by the degree distribution $P_k$ when the joint degree-degree distribution is $P(k,k') = k P_k k' P_{k'} / z^2$, where $z = \sum_{k=0}^\kmax k P_k$ is the mean degree and $\kmax$ is the maximum degree in the network.  We assume that the network topology is locally tree-like (which implies that the number of short loops is small).\cite{Melnik11}

We start by calculating the fraction $\rho_k^{(i)}(n)$ of degree-$k$ nodes that are $S_i$-active ($i \in \{1,2\}$) at the $n$th time step of the synchronous update process (i.e., after the $n$th synchronous update of all nodes).  We thus consider a randomly chosen degree-$k$ node $A$ and calculate its probability of being $S_i$-active at time step $n$.  We choose $A$ uniformly at random, so this probability is $\rho_k^{(i)}(n)$.  As an initial condition, we set a fraction $\phi^{(i)}$ of all nodes to be $S_i$-active.  Thus, node $A$ is initially $S_i$-active with probability $\phi^{(i)}$.  If it is not initially $S_i$-active, then (as discussed in the main text) it can become $S_i$-active after a synchronous update with probability $F_i(m_1,m_2,k)$.  The arguments $m_1$ and $m_2$ are, respectively, the numbers of $A$'s neighbors in states \sone~and \stwo~before the update.


We denote by $\bar q_k^{(i)}(n-1)$ the probability that at time step $n-1$ (i.e., immediately before the $n$th update of node $A$) a random neighbor of node $A$ is $S_i$-active, conditioned on node $A$ itself not being $S_i$-active.~\cite{Note3}  Thus, the probability that exactly $m_1$ of the $k$ neighbors of node $A$ are \sone-active is given by $B^k_{m_1}\lb(\bar q_k^{(1)}(n-1)\rb)$, where
\begin{align}
	B_m^k(q)=\binom{k}{m} q^m (1-q)^{k-m}
\end{align}
is the binomial distribution.  Similarly, the probability that exactly $m_2$ of these $m_1$ \sone-active neighbors are also \stwo-active is given by $B^{m_1}_{m_2}\lb(\bar q_k^{(2)}(n-1)\left/\bar q_k^{(1)}(n-1)\rb)\right.$.  Note that $\bar q_k^{(2)}(n-1)\left/\bar q_k^{(1)}(n-1)\right.$ is the probability that an \sone-active neighbor of node $A$ is also \stwo-active.  Combining these probabilities yields (for $i=\{1,2\}$)
\begin{align} \label{rho_i}
	\nn \rho_k^{(i)}(n) =& \phi^{(i)} +  \lb(1-\phi^{(i)}\rb) \sum_{m_1=0}^k  B_{m_1}^{k}\lb(\bar q_k^{(1)}(n-1)\rb) \\
	&\times \sum_{m_2=0}^{m_1} B_{m_2}^{m_1}\lb(\frac{\bar q_k^{(2)}(n-1)}{\bar q_k^{(1)}(n-1)}\rb)F_{i}(m_1,m_2,k)\,.
\end{align}
In Eq.~\eqref{rho_i}, we have assumed that the states of any two neighbors of node $A$ are independent.  We would expect this to be the case for a graph that is locally tree-like, such as random networks constructed using the configuration model.  Although this assumption breaks down on real-world networks with high clustering coefficients and/or significant community structure, it has been demonstrated recently using several dynamical processes (including single-stage complex contagions) that cascade results obtained using locally tree-like approximations often remain valid.~\cite{Melnik11}

\begin{figure}[h]
\centering
\includegraphics[width=0.9\columnwidth]{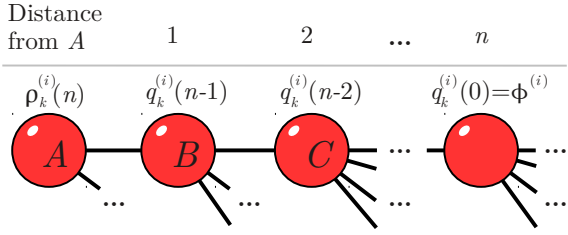}
\caption{Tree-like structure of a network near node $A$, which we treat as the root of the tree. For every two nodes connected by an edge (e.g., nodes $B$ and $C$), the node that is closer to $A$ is called the parent; thus, node $B$ is the parent of node $C$, and node $C$ is the child of node $B$. Only the influence from nodes within a distance $n$ of node $A$ can reach $A$ in $n$ time steps.}
\label{tree}
\end{figure}

A neighbor $B$ of the degree-$k$ node $A$ has degree $k'$ with probability $P(k,k')\left/\sum_{k'} P(k,k')\right.$. Therefore, assuming that $A$ is not $S_i$-active, we can express $\bar q_k^{(i)}$, the probability that a random neighbor of $A$ is $S_i$-active in terms of $q_{k'}^{(i)}$, defined as the probability that a degree-$k'$ neighbor of $A$ is $S_i$-active~\cite{Note4}.  The formula is
\begin{align} \label{qbar}
 	\bar q_k^{(i)}(n) = \frac{\sum_{k'} P(k,k') q_{k'}^{(i)}(n)}{\sum_{k'} P(k,k')}\,.
\end{align}
To calculate $q_{k}^{(i)}$ and thus $\bar q_{k}^{(i)}$, we will establish a recurrence relation for $q_{k}^{(i)}(n)$ and use the fact that $q_{k}^{(i)}(0) = \phi^{(i)}$ (i.e., using the fact that all nodes are initially $S_i$-active with probability $\phi^{(i)}$). Consider node $B$, which is a neighbor of node $A$. Using similar reasoning as for Eq.~\eqref{rho_i}, we express the probabilities $q_{k}^{(i)}$ that $B$ is $S_i$-active, given that $A$ is not $S_i$-active, in terms of probabilities $\bar q_{k}^{(i)}$. The probabilities $\bar q_{k}^{(i)}$ are the probabilities that $B$'s children (i.e., neighbors of $B$ that are one step further away from $A$; see Fig.~\ref{tree}) are $S_i$-active at the previous time step, given that $B$ is not $S_i$-active:
\begin{align} \label{q1}
	\nn q_k^{(1)}(n+1) = &\phi^{(1)} +  \lb(1-{\phi^{(1)}}\rb) \sum_{m_1=0}^{k-1} B_{m_1}^{k-1}\lb(\bar q_k^{(1)}(n)\rb)\\
 \times & \sum_{m_2=0}^{m_1} B_{m_2}^{m_1}\lb(\frac{\bar q_k^{(2)}(n)}{\bar q_k^{(1)}(n)}\rb) F_1(m_1,m_2,k)
\end{align}
and
\begin{align} \label{q2}
	\nn q_k^{(2)}(n+1) = &\phi^{(2)} +  \lb(1-{\phi^{(2)}}\rb) \sum_{m_1=0}^{k-1} B_{m_1}^{k-1}\lb(\bar q_k^{(1)}(n)\rb) \\
\nn \times &\sum_{m_2=0}^{m_1} B_{m_2}^{m_1}\lb(\frac{\bar q_k^{(2)}(n)}{\bar q_k^{(1)}(n)}\rb) \lb[ \lb(1-\bar q_k^{(1)}(n)\rb) \rb. \\
	\times&  F_2(m_1,m_2,k) + \lb. \bar q_k^{(1)}(n) F_2(m_1+1,m_2,k)\rb]\,.
\end{align}
By the definition of $q_{k}^{(1)}$, the parent $A$ of node $B$ is not \sone-active. Therefore, unlike Eq.~\eqref{rho_i}, the sum over $m_1$ in Eq.~\eqref{q1} runs to $m_1=k-1$, which implies that $B$ can only be \sone-activated by its $k-1$ children if it is not part of the seed. Similarly, $A$ is not \stwo-active when calculating $q_{k}^{(2)}$ in Eq.~\eqref{q2}. However, even though $A$ is not \stwo-active, it could still be \sone-active and thereby contribute to \stwo-activation of $B$. We take this into account in the quantity in square brackets in Eq.~\eqref{q2}: with probability $\bar q_k^{(1)}$, the parent of $B$ is \sone-active, so the number $m_1$ of \sone-active neighbors of $B$ is increased by $1$ in $F_2$; meanwhile, $m_1$ in $F_2$ is unchanged with complementary probability $1-\bar q_k^{(1)}$.

One can write Eqs.~\eqref{rho_i}--\eqref{q2} in concise form using vector notation.  Writing $Q^{(i)}(n) = \lb[q_{1}^{(i)}(n),\ldots, q_\kmax^{(i)}(n)\rb]$ yields
\begin{align}
	\label{Rho} \rho_k^{(i)}(n) &= h_k^{(i)}\lb( Q^{(1)}(n-1), Q^{(2)}(n-1) \rb)\,,\\
	\label{Q} Q^{(i)}(n) &= g^{(i)}\lb( Q^{(1)}(n-1), Q^{(2)}(n-1) \rb)\,.
\end{align}
Starting with $Q^{(i)}(0) =\lb[\phi^{(i)},\ldots,\phi^{(i)}\rb]$, we iterate Eqs.~\eqref{Rho}--\eqref{Q} to obtain $\rho_k^{(i)}(n)$ for all $n$ and $k$.  Because node $A$ has degree $k$ with probability $P_k$, the aggregate fraction of $S_i$-active nodes at time step $n$ is given by
\begin{align}
	 \rho^{(i)}(n) = \sum_k P_k \rho_k^{(i)}(n)\,.
\end{align}

Additionally, note that Eqs.~\eqref{rho_i}--\eqref{q2} can be simplified for configuration-model networks, as the degrees of nodes at the two ends of an edge are independent, so the joint degree-degree distribution factorizes as $P(k,k') = k P_k k' P_{k'} / z^2$. In this case, it follows from Eq.~\eqref{qbar} that $\bar q^{(i)}=\sum_k k P_k q^{(i)}_k/z$ is degree-independent. One thereby obtains
\begin{align}\label{rho_pk}
	\nn \rho^{(i)}(n+1) &= {\phi^{(i)}} +  \lb(1-\phi^{(i)}\rb) \sum_k P_k \sum_{m_1=0}^k  B_{m_1}^{k}\lb(\bar q^{(1)}(n)\rb) \\
	&\times \sum_{m_2=0}^{m_1} B_{m_2}^{m_1}\lb(\frac{\bar q^{(2)}(n)}{\bar q^{(1)}(n)}\rb)F_{i}(m_1,m_2,k),
\end{align}
\begin{align}\label{q1_pk}
	\nn \bar q^{(1)}(n+1) &= \phi^{(1)} +  \lb(1-{\phi^{(1)}}\rb) \sum_k \frac{k P_k}{z} \\
	\nn &\times \sum_{m_1=0}^{k-1} B_{m_1}^{k-1}\lb(\bar q^{(1)}(n)\rb)\sum_{m_2=0}^{m_1} B_{m_2}^{m_1}\lb(\frac{\bar q^{(2)}(n)}{\bar q^{(1)}(n)}\rb) \\ &\times  F_1(m_1,m_2,k),
\end{align}
\begin{align}\label{q2_pk}
	\nn \bar q^{(2)}(n+1) &= \phi^{(2)} +  \lb(1-{\phi^{(2)}}\rb) \sum_k \frac{k P_k}{z} \\
	\nn &\times \sum_{m_1=0}^{k-1} B_{m_1}^{k-1}\lb(\bar q^{(1)}(n)\rb) \sum_{m_2=0}^{m_1} B_{m_2}^{m_1}\lb(\frac{\bar q^{(2)}(n)}{\bar q^{(1)}(n)}\rb) \\
	\nn &\times \lb[ \lb(1-\bar q^{(1)}(n)\rb) F_2(m_1,m_2,k) \rb. \\
	&+ \lb. \bar q^{(1)}(n) F_2(m_1+1,m_2,k)\rb].
\end{align}
Note, however, that Eqs.~\eqref{rho_pk}--\eqref{q2_pk} are inaccurate for networks with degree-degree correlations.  For example, they fail to predict the cascades illustrated in Figs.~\ref{S2drS1a_asynch} and~\ref{S1drS2a_asynch}.

We now consider asynchronous updating, in which we update only a fraction $\tau$ of nodes at each time step.  We choose the time step $\D t=\tau$ to have a common time scale for all $\tau$ (including the synchronous updating case of $\tau=1$).  If the updating is synchronous (i.e., if $\tau=1$), then the probability $Q^{(i)}$ increases by $\D Q^{(i)} = g^{(i)}\lb( Q^{(1)},Q^{(2)} \rb) - Q^{(i)}$.  In other words, all nodes that are available for activation are activated.  In the asynchronous updating case, $\D Q^{(i)} = \tau\lb(g^{(i)}\lb( Q^{(1)},Q^{(2)} \rb) - Q^{(i)}\rb)$.  Therefore, for sufficiently small values of $\tau$, the temporal evolutions of $Q^{(i)}$ and $\rho_k^{(i)}$ can be approximated as continuous.  This yields the following set of ordinary differential equations:
\begin{align}
	\dot \rho_k^{(i)}(t) &= h_k^{(i)}\lb( Q^{(1)}(t), Q^{(2)}(t) \rb)- \rho_k^{(i)}(t)\\ \nn &\equiv H_k^{(i)}\lb(\rho_k^{(i)}(t),Q^{(1)}(t),Q^{(2)}(t)\rb)\,,\\
	\dot Q^{(i)}(t) &= g^{(i)}\lb( Q^{(1)}(t),Q^{(2)}(t) \rb) - Q^{(i)}(t) \\ \nn &\equiv G^{(i)}\lb(Q^{(1)}(t),Q^{(2)}(t)\rb)\,,
\end{align}
which are Eqs.~\eqref{rhoki}--\eqref{qki} of the main text.


\section{Additional Features of the Model} \label{appB}

Our multi-stage model possesses several additional interesting features that we did not discuss in the main text.  They motivate the development of more accurate analytical approximations than the one that we presented in this paper.  The reason for this is that the above theory, which we generalized from previous work on single-stage models and other dynamical systems,~\cite{Gleeson07a,Gleeson08a} neglects some effects that arise in the multi-stage model.  These features are either minimal or absent entirely from single-stage contagions models, so the need to develop more accurate analytical approximations has become apparent only because of the new model in this paper.  Below we highlight two features that we have identified as requiring more accurate modelling.

The first feature, which amounts to believing one's own gossip, is similar to the ``June bug effect'' from sociology,~\cite{junebug} except that here the feedback is of primary importance.  This feature can be understood by considering a node that is initially \sone-active but not \stwo-active.  Suppose that this node \sone-activates some of its neighbors at a subsequent time step.  After that, these \sone-active neighbors can in turn \stwo-activate the original node.  That is, a node can become more active because of the feedback peer pressure that it experiences from the neighbors it had \sone-activated in the first place.  The analytical approximation that we have employed does not account for this effect, as it assumes that neighbors activated by a node are distributed across the network (rather than gathered around the node).  The ensuing theory thus can underestimate the number of \stwo-active nodes in the system.

\def\fwidth{0.99}
\begin{figure}[h]
\centering
\includegraphics[width=\fwidth\columnwidth]{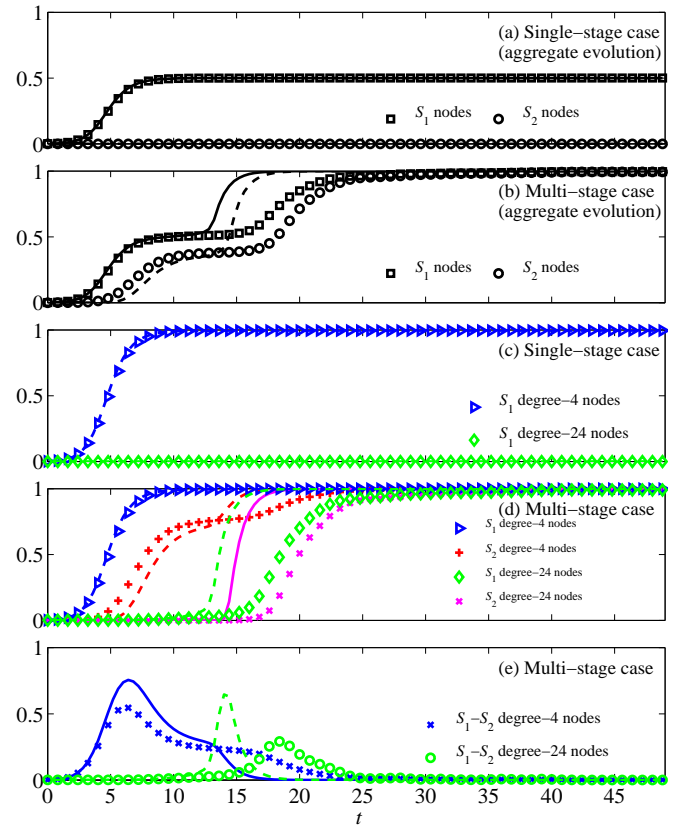}
\caption{This figure is analogous to Fig.~\ref{S2drS1a_asynch} of the main text, but here we use a different value of the upper threshold $R_2$ for the multi-stage case [panels (b), (d), and (e)].  In this example, we use the threshold value $R_2=0.8$; in Fig.~\ref{S2drS1a_asynch}, we used $R_2=0.7$.  As discussed in the text, the state-segregation effect is more pronounced for $R_2=0.8$, which implies that all of the neighbors of degree-4 nodes must be \sone-active for such a node to become \stwo-active (whereas only three \sone-active neighbors are needed for such \stwo-activation when $R_2=0.7$).  Our analytical approximation does not properly account for state segregation, and the consequences of that can be seen by comparing panels (d) in the two figures. The match with the numerical simulations in this figure is clearly worse than that in Fig.~\ref{S2drS1a_asynch}.
}
\label{S2drS1a_asynch_R08}
\end{figure}

The second feature, which we call ``state segregation'', leads to theoretical predictions that overestimate the numerically observed fractions of active nodes [see Fig.~\ref{S2drS1a_asynch_R08}(d)].  In this figure, we consider (4,24)-correlated random networks, and the theory predicts that \sone-activation of degree-24 nodes happens earlier than what we observe numerically.  This arises as follows.  Degree-24 nodes begin to become \sone-active because they are activated by their \stwo-active degree-4 neighbors.  However, degree-4 nodes first become \stwo-active [for the threshold value $R_2=0.8$ used in Fig.~\ref{S2drS1a_asynch_R08}(d)] only if they are connected exclusively to other degree-4 nodes who are all \sone-active at that time.  Degree-4 nodes do not experience sufficient peer pressure to become \stwo-active if they are connected to any of the degree-24 nodes (who are all inactive at that time).  In other words, there is a negative correlation between a degree-4 node becoming \stwo-active and that node being connected to an inactive degree-24 node.  Our theory does not take this anticorrelation into account; it assumes that inactive degree-24 nodes are equally likely to have connections to degree-4 nodes of any state.  This assumption does not hold when the first \stwo-activations of degree-4 nodes occur.  Therefore, the theory assumes that there are at least some connections between inactive degree-24 nodes and early \stwo-active degree-4 nodes, even though there are no such connections in the network.  As a result, the theory overestimates the fraction of \sone-active degree-24 nodes in Fig.~\ref{S2drS1a_asynch_R08}(d).

The consequences of state segregation become less pronounced when the \stwo-threshold is reduced from $R_2=0.8$ to $R_2=0.7$.  The degree-4 nodes now become \stwo-active if at least 3 (rather than 4) of their neighbors are \sone-active, thereby relaxing the restriction of not being connected to any of the degree-24 nodes.  One can see this in Fig.~\ref{S2drS1a_asynch}(d) in the main text, where the agreement between theory and numerical results is clearly better than in Fig.~\ref{S2drS1a_asynch_R08}(d).

It is also important to recognize that our multi-stage model requires the use of a more comprehensive set of probabilities than the probabilities $\bar q_k^{(i)}(n)$ defined in Appendix~\ref{appA}.  For example, strictly speaking $\bar q^{(1)}_k$ can be used in Eq.~\eqref{q2} for calculating $q^{(2)}_k$ only in the case of a direct transition of the parent node from state \szero~to state \stwo.  The correctness of the use of $\bar q^{(1)}_k$ for calculating $q^{(2)}_k$ depends on the extent to which the fact that the parent node is \sone-active will affect the probability that it will become \stwo-active.  In many cases, the use of $\bar q^{(1)}_k$ is adequate.  In general, however, if the parent node is already \sone-active, then the probability that its child is \sone-active will be higher than what is given by $\bar q^{(1)}_k$.  For this reason, the analytical approximation given in Appendix~\ref{appA} can underestimate (or even fail to predict) the cascades of activations in some parameter regimes.  It is therefore desirable to develop better (but more complicated) approximations.


\section{Cascade Condition and Bifurcation Analysis} \label{appC}

We now derive the \emph{cascade condition} and describe the computational bifurcation analysis that we illustrated in Fig.~\ref{figcc} in the main text. The cascade condition determines whether an infinitesimally small seed fraction of activated nodes triggers a finite-size cascade (i.e., a cascade in which the number of active nodes scales linearly with the size of the network).  We present a derivation for uncorrelated random networks with degree distribution $P_k$.~\cite{Note5}  It is sufficient to analyze the dynamics of the auxiliary variables $\bar q^{(i)}$, which are described by Eqs.~\eqref{q1_pk} and~\eqref{q2_pk}.  To avoid restating lengthy formulas, we rewrite these equations as
\begin{equation}
	q^{(i)}=\tilde{g}^{(i)}\left(q^{(1)},q^{(2)}\right)\,,
\end{equation}
where we have dropped the bars on $\bar q^{(i)}$ to avoid cluttering the notation. The dynamics under asynchronous updating can be approximated by the ordinary differential equations
\begin{align}
	\frac{dq^i}{dt} &=\tilde{g}^{(i)}\left(q^{(1)},q^{(2)}\right)-q^{(i)} \equiv \tilde{G}^{(i)}\left(q^{(1)},q^{(2)}\right)\,,
\end{align}
which, written in vector form, results in the two-dimensional dynamical system
\begin{equation}
	\frac{dq}{dt}=\tilde{g}(q)-q\,.
\end{equation}
Because the equilibria for both synchronous and asynchronous updating are the same, our cascade condition and bifurcation analysis are valid for both cases. The function $\tilde{g}$ is monotone increasing, so each of the partial derivatives of $\tilde{g}^{(i)}$ with respect to $q^{(j)}$ is positive. That is, ${\rm D}_{j}\tilde{g}^{(i)}>0$. Hence, one finds (e.g., by direct calculation) that the eigenvalues of the Jacobian matrix ${\rm D}\tilde{g}(q)$ [and consequently the eigenvalues of ${\rm D}\tilde{G}(q)$] are real.

In analogy to the methodology proposed for single-stage cascades,~\cite{Gleeson07a,Gleeson08a} a small initial seed will grow if one of the (real) eigenvalues of ${\rm D}\tilde{G}(0)$ is positive. This can be established by checking the sign of the determinant of ${\rm D}\tilde{G}(0)$, which yields the following condition for global cascades:
\begin{equation}
	{\rm D}_{2}\tilde{g}^{(1)}_{0}{\rm D}_{1}\tilde{g}^{(2)}_{0}-\left({\rm D}_{1}\tilde{g}^{(1)}_{0}-1\right)\left({\rm D}_{2}\tilde{g}^{(2)}_{0}-1\right)>0\,,
\label{ccond}
\end{equation}
where ${\rm D}_{j}\tilde{g}^{(i)}_{0}$ denotes the partial derivative of $\tilde{g}^{(i)}$ with respect to $q^{(j)}$ evaluated at $q=0$.  The partial derivatives are given by
\begin{align*}
	{\rm D}_{1}\tilde{g}^{(1)}_{0}=&\left(1-\phi^{(1)}\right)\sum_{k}\frac{k(k-1)P_k}{z}\\
	&\times \left[F_1(1,0,k)-F_1(0,0,k)\right]\,,\\
	{\rm D}_{2}\tilde{g}^{(1)}_{0}=&\left(1-\phi^{(1)}\right)\sum_{k}\frac{k(k-1)P_k}{z}\\
	&\times \left[F_1(1,1,k)-F_1(1,0,k)\right]\,,\\
	{\rm D}_{1}\tilde{g}^{(2)}_{0}=&\left(1-\phi^{(2)}\right)\sum_{k}\frac{k^2P_k}{z}\left[F_2(1,0,k)-F_2(0,0,k)\right]\,,\\
	{\rm D}_{2}\tilde{g}^{(2)}_{0}=&\left(1-\phi^{(2)}\right)\sum_{k}\frac{k(k-1)P_k}{z}\\
	&\times \left[F_2(1,1,k)-F_2(1,0,k)\right]\,.
\end{align*}
Note that if $\beta=0$, then $F_i(1,1,k)=F_i(1,0,k)$.  Therefore, ${\rm D}_{2}\tilde{g}^{(1)}_{0}={\rm D}_{2}\tilde{g}^{(2)}_{0}=0$ and Eq.~\eqref{ccond} reduces to the single-stage cascade condition derived in Ref.~\onlinecite{Gleeson07a}.  For networks with degree-degree correlations, the Jacobian matrix has a much higher dimension, so the eigenvalues must typically be located numerically.

Although the derivation of the cascade condition results in a closed-form expression (\ref{ccond}), it is a fairly crude approximation.  This can be seen in Fig.~\ref{figcc}, where the area of parameter space in which cascades occur is larger than that predicted by the cascade condition.  We now show that this discrepancy arises in part from the value at which the partial derivatives in~\eqref{ccond} are evaluated. The fact that the eigenvalues of ${\rm D}\tilde{g}(q)$ [and hence of ${\rm D}\tilde{G}(q)$] are real rules out the possibility of Hopf bifurcations.  It follows that typical bifurcations will be of saddle-node type, since the lack of symmetry means that other types of local, codimension 1 bifurcations must be non-generic. One can locate such bifurcations accurately using linear stability analysis and numerical computations.  Toward this end, we define the small perturbation $\xi=q-q_*$ and linearize about the equilibrium point $q_*$ to obtain
\begin{equation}
	\dot{\xi}=\left[{\rm D}\tilde{g}(q_*)-I\right]\xi\,,
\end{equation}
where ${\rm D}\tilde{g}(q_*)$ is the Jacobian matrix of $\tilde{g}$ evaluated at the equilibrium point $q_*$ (so its components are ${\rm D}_{j}\tilde{g}^{(i)}_{*}$) and $I$ is the identity matrix.  The equilibrium point $q_*$ is unstable when one of the eigenvalues of ${\rm D}\tilde{g}(q_*)-I$ is positive.  In contrast to the derivation of the cascade condition, in which we set $q_*=0$, one must now determine the equilibrium point $q_*$, which is typically only possible via numerical computation.  To locate the bifurcation, one must solve
\begin{equation}
	q^{(i)}_*=\tilde{g}^{(i)}_*\left(q^{(1)}_*,q^{(2)}_*\right)\,, \quad i \in \{1,2\}
\label{equilibria}
\end{equation}
for the equilibria and
\begin{equation}
	{\rm D}_{2}\tilde{g}^{(1)}_{*}{\rm D}_{1}\tilde{g}^{(2)}_{*}-\left({\rm D}_{1}\tilde{g}^{(1)}_{*}-1\right)\left({\rm D}_{2}\tilde{g}^{(2)}_{*}-1\right)=0\,
\label{eig}
\end{equation}
for the zero eigenvalue. (Note that~\eqref{eig} is similar to the cascade condition~\eqref{ccond}, but the partial derivatives in the former are evaluated at the equilibrium point $q=q_*$.) Using one model parameter as a free variable (with all other parameters held constant), one can solve equations~\eqref{equilibria} and~\eqref{eig} numerically and thereby determine the location of the saddle-node bifurcation. It is then possible to use numerical continuation to trace the bifurcations as a second parameter is varied.  We show the results of this continuation in Fig.~\ref{figcc} of the main text.


\begin{thebibliography}{10}
\bibitem{notehipster}%
  \BibitemOpen
  \bibinfo {note} {However, assumption (i) should probably be relaxed for a model that includes ``hipsters''.}
  \BibitemShut{Stop}%
\end{thebibliography}
\end{document}